# Bayesian evidence for Finite element model updating

Linda Mthembu, PhD student, Electrical and Information Engineering, University of the Witwatersrand, Johannesburg Private Bag 3, Johannesburg, 2050, South Africa
Tshilidzi Marwala, Professor of Electrical and Information Engineering, University of the Witwatersrand, Private Bag 3, Johannesburg, 2050, South Africa
Michael I. Friswell, Sir George White Professor of Aerospace Engineering, Department of Aerospace Engineering, Queens Building, University of Bristol, Bristol BS8 1TR, UK
Sondipon Adhikari, Chair of Aerospace Engineering, School of Engineering, Swansea University, Singleton Park, Swansea SA2 8PP, United Kingdom

**Abstract**

This paper considers the problem of *model selection* within the context of finite element model updating. Given that a number of FEM updating models, with different updating parameters, can be designed, this paper proposes using the Bayesian evidence statistic to assess the probability of each updating model. This makes it possible then to evaluate the need for alternative updating parameters in the updating of the initial FE model. The model evidences are compared using the Bayes factor, which is the ratio of evidences. The Jeffrey's scale is used to determine the differences in the models. The Bayesian evidence is calculated by integrating the likelihood of the data given the model and its parameters over the *a priori* model parameter space using the new nested sampling algorithm. The nested algorithm samples this likelihood distribution by using a hard likelihood-value constraint on the sampling region while providing the posterior samples of the updating model parameters as a by-product. This method is used to calculate the evidence of a number of plausible finite element models.

**Nomenclature**

| | |
|---|---|
| $\theta$ | Model Parameters |
| $D$ | Real measured system data |
| $H$ | Finite element model (Mathematical model) |
| $Z$ | Evidence |
| $N$ | Number of samples |
| $Max$ | Maximum number of iterations |
| $L$ | Likelihood probability |
| $\pi$ | Prior Probability |
| $c$ | Damping matrix |
| $k$ | Stiffness matrix |
| $m$ | Mass matrix |
| $\ddot{x}(t)$ | Node acceleration |
| $\dot{x}(t)$ | Node velocity |
| $x(t)$ | Node displacement |
| $w$ | Natural frequency vector |
| $\phi$ | Mode shape vector |
| FEM | Finite element model |
| FEMUP | Finite element model updating problem |
| PDF | Probability Distribution Function |
| MCMC | Markov Chain Monte Carlo |

# 1. Introduction

System modeling forms an important stage of many engineering design problems. The results from the model either confirm or highlight limitations of the design. An analyst is usually interested in the accuracy, confidence range and more critically the correctness of the assumed mathematical model. In this paper the model domain is structural finite element models (FEMs). These models are used to approximate the dynamics of structural systems, e.g. train chassis, aircraft fuselages, bicycle frames, civil structures etc. The finite element model updating problem (FEMUP) arises when a real system's dynamic behavior is measured (e.g. the natural frequencies at which particular system deformations occur) and the results of the mathematical model of that system do not correspond to the measured data [6, 8]. This problem is compounded by the fact that a multitude of mathematical models of the structure, with varying levels of complexity, can be developed, leading to non-unique solutions for a particular system.

Finite element models are limited by definition; they are an *approximation* of a real system and will thus never produce dynamic results that are *equal* to the measured system's data. The challenge is then what can be done to the initial model for it to better reflect the real system's dynamic results? This leads to the need for intelligently improved or updated models. Specifically an automatic methodology of determining salient model parameters that require updating needs to be developed. This has to be attained whilst using realistic characteristic parameters of the system in question. Two main directions of research have been established in the area of finite element model updating (FEMU); direct and indirect (iterative) methods [8].

In the direct model updating paradigm [3, 5, 8] the model modal parameters are directly equated to the measured modal data. Model updating is then characterized by the direct updating of mass or stiffness matrix elements. This effectively constrains the modal properties and frees the system matrices for updating. This approach often results in unrealistic elements in the system matrices e.g. large and physically impossible mass elements. In the indirect or iterative model updating approach the updating problem is formulated as a 'relaxed' optimization problem, often approached by the use of maximum likelihood methods [8, 9, 12 and 19]. A select few parameters are varied in model *elements* and the resulting model is optimized to minimize its difference from the measured data. Given the non-uniqueness of updating models, this paper proposes going back to the initial stage of modeling and questions the choice of the initial finite element model(s). We propose the FEMUP should initially be approached from a *model selection* [2, 10, 17] perspective.

Before model updating, a number of fundamental questions need to be addressed, these are; (i) Is the model we want to update the correct one? (ii) Which aspects of the model do we need to update? (iii) Having determined the updating parameters, with what certainty can we guarantee that our updated model is correct? The first point can be recast as; what *evidence* do we have that our model is the correct one given that a number of models can be generated? This we propose ought to be an essential and necessary statistic to establish before any model updating scheme can proceed. Having established that the most probable model is found from a set of possible models we can then focus on the updating process. In this paper we focus on the first problem which lends itself well to Bayesian inference. Bayesian analysis allows one to deal with initial and subsequent finite element model uncertainties in an intuitive and systematic way [2, 10]. This paper is concerned with calculating the evidence of finite element models.

In the next section we formally present the finite element model formulation and very compactly review the current approaches to the FEMUP. In section 3 we review Bayes theorem and introduce the evidence calculation in the Bayesian framework. Section 4 presents the evidence calculation algorithm. Section 5 presents proof of concept simulations of evidence calculation using a simple beam structure. We then conclude the paper.

## 2. Finite element background

2.1 Formal definition

In engineering, dynamic structures are often analyzed from bottom to systems (component) level. At the bottom most level the structure is discretized into constituent elements, for mathematical analysis and computational feasibility, to a system of a second-order matrix of the form [6, 8]:

$$M\ddot{x}(t) + C\dot{x}(t) + Kx(t) = 0 \qquad (1)$$

where $M$, $C$ and $K$ are of equal size and are called the mass, damping and stiffness matrices respectively or system matrices. Assuming each finite element's dynamic displacement response behaves according to $x(t) = \phi e^{-wt}$ equation 1 above is transformed to:

$$[-w^2 M + wC + K]\phi = 0 \qquad (2)$$

In cases where the structure is small, lightly damped or undamped the damping matrix $C = 0$ the above equation reduces to:

$$[-w_j^2 M + K]\phi_j = \{\varepsilon_j\} \qquad (3)$$

where $w_j$ and $\phi_j$ are the j$^{th}$ system natural frequency and mode shapes, together known as modal properties, $\{\varepsilon_j\}$ is the j$^{th}$ error of the error vector. Given a set of measured real system modal data, the FEMU problem is then for the model to realistically approximate the mass and stiffness matrices that will produce modal data that is as close to the real systems' as possible. If the model does not result in matching modal properties the error vector is non-zero and some model parameters will need to be updated. Basic physics relates elastic element stiffness with Young's modulus; the mass is a function of its geometry and density. By considering these variables as random values that are defined within certain intervals for particular materials, it is possible to closely approximate the measured modal values in a formal way. In this paper we treat the Young's modulus as a variable to be updated. Measurement of structural dynamics produces one set of, not necessarily repeatable data, e.g. building vibrations, earthquakes and vehicle dynamic performances. Traditional approaches to prevent data biasing models are to repeat the measurements a number of times to generate a general trend and averaging the data. This is not a simple process and can become very expensive [6]. This situation is well suited to the Bayesian approach as opposed to a frequentist paradigm. In Bayesian inference we are given one observed data set which our model is supposed to approximate well. In the frequentist approach the data acquisition or experiment is assumed to be repeatable, such that a pattern could be determined from a large dataset. The Bayesian perspective is bold in that it infers a lot from one dataset [2]. In the next section we present the tools of this paradigm, Bayes theorem and then explain Bayesian inference in the context of the finite element updating problem.

## 3. Bayesian Inference

In current times of massive data, the ability to quantify the model's posterior confidence and to compare one model's ability to approximate data with another has become the focus of data analysis. One paradigm in this direction that has seen a recent resurgence is Bayesian inference [2, 10 and 20]. Bayesian inference allows one to quantify uncertainties in quantities of interest in a formal way. Bayesian inference is often implemented in two settings; parameter estimation and model selection. Parameter estimation is concerned with the plausibility of a given model's parameters and this is often carried out using standard sampling methods e.g. Markov Chain Monte Carlo (MCMC) [2, 10, 11 and see 16 for recent advances in these techniques]. Model selection on the other hand deals with the evidence for each candidate mathematical model to approximate a particular observed dataset.

3.1 Parameter estimation

In parameter estimation the mathematical model is assumed to be true, the model is then 'fitted' to the data and the posterior plausibility of the model parameters can be inferred. This probability is calculated via Bayes theorem as follows:

$$P(\theta \mid D, H) = \frac{P(D \mid \theta, H) P(\theta \mid H)}{P(D \mid H)} \qquad (4)$$

where the left hand side is the posterior probability of the updating parameters for the true model H, given some data D, the prior probability of the model parameters is $P(\theta \mid H)$, $P(D \mid \theta, H)$ is called the likelihood of the model. The denominator $P(D \mid H)$ is called the marginal likelihood, or the *evidence*, of the model where the parameters have been marginalized out. Bayes theorem automatically incorporates the updating of the parameters by definition; it updates the a priori probability distribution of the model parameters with the likelihood of the model explaining the measured data. In the FEM context the parameter variables that typically govern the dynamics of the finite elements of the mathematical model are the following: Young's modulus, cross-sectional areas, damping coefficient etc. These parameters in turn affect the stiffness, mass and damping matrices in equation 1. In [11] the *parameter estimation* context of the Bayesian framework was implemented to obtain the posterior probability distribution of model parameters. By obtaining the posterior probability of the updating parameters the probability distribution of the modal properties may then be calculated. This gives the researcher a quantitative measure of the probable ranges of the modal and updating parameters of the model.

In this paper the interest is not in the posterior probability of the updating parameters per se but in the evidence of the chosen updating model thus a *model selection* problem although the algorithm implemented provides the posterior probabilities of the updating parameters as a by-product.

3.2 Model Comparison

In model comparison one dataset is observed and a number of possible models can be formulated. Bayesian inference provides a platform for calculating which model is the most plausible from this set. The posterior probability of each model within a set of plausible models is given by Bayes theorem:

$$P(H_i | D) = \frac{P(D | H_i) P(H_i)}{P(D)} \quad (5)$$

where $P(H_i)$ is the prior probability of each model and $P(D)$ is the probability of the data. Since for all models the denominator is independent of the model we may ignore it and the theorem then reduces to:

$$P(H_i | D) \propto P(D | H_i) P(H_i) \quad (6)$$

The first term on the right of equation 6 is the evidence from equation 4. Given that each model may be equally likely to fit the data, the evidence term is the deciding factor on which model is the most plausible for a particular observed dataset. One standard method of comparing models in Bayesian analysis is Bayes factor defined as;

$$Bayes\,Factor = \frac{P(D|H_i)}{P(D|H_j)} = \frac{\int d\theta_i\, P(D|\theta_i, H_i) P(\theta_i | H_i)}{\int d\theta_j\, P(D|\theta_j, H_j) P(\theta_j | H_j)} \quad (7)$$

So by calculating each model's evidence we can quantify their ratios. In [1, 4 and 17] the concept of model selection for engineering structures was used. In [1] the posterior distribution of the model was assumed to be Gaussian which only works for certain types of models [1, 2, 4, 10 and 17]. The method proposed in this paper does not place any assumption of the form of the posterior distribution of the model. In [17] a recently proposed MCMC type posterior probability sampling algorithm (TMCMC from [4]) was implemented. This algorithm estimates the model evidence by sampling the posterior probability distribution of the model by a sequence of non-normalized intermediate probability functions. This algorithm suffers from the use of many free parameters; e.g. a variable to balance the sampling steps, the number of intermediate probability distribution functions (PDF), the tempering parameter and the sample weights [4, 17]. The algorithm proposed here is believed to be simpler and to have fewer free-parameters.

In the next section we explain the Bayesian evidence. We then introduce an algorithm called nested sampling for efficiently estimating the model evidence that is theoretically and practically simpler than TMCMC.

3.3 Bayesian evidence

From the previous equation each model's evidence is given by equation 8:

$$evidence = P(D | H_i) = \int d\theta\, P(D | \theta, H_i) P(\theta | H_i) \quad (8)$$

This equation may be interpreted as follows; given a unit of parameter space $d\theta$ with a model having a prior probability of $P(\theta|H_i)$ over this space, the posterior probability distribution of the model over the same space will depend on how the model parameters fit the data $P(D|\theta,H_i)$. In this paper the likelihood function is defined as the normalized exponential error between measured and real natural frequencies (the error given in equation 1) at the measured modes as in [Ref.13]. The model posterior probability will be highest at the most probable parameter set which occupies a small region of the original parameter space. In simple models an initially smaller, by definition, prior parameter space would almost be fully utilized to fit the data as the posterior probability would occupy a large portion of the prior parameter space resulting in a correspondingly large evidence value. Complex models have larger parameter spaces because they have many free parameters that allow them to fit almost any data and this often results in over-fitting [2, 10]. This larger parameters space is 'wasted' in accounting for the resultant, relatively narrow posterior density. The Bayesian inference paradigm thus automatically penalizes complex models without the need for a model regularization term often incorporated in non-Bayesian approaches [2, 10, 19].

Continuing with the evidence formulation, equation 8 can be re-written in the following form:

$$Z = \int L(\theta)\,\pi(\theta)\,d\theta \tag{9}$$

Analytically evaluating this integral may be difficult or impossible if the product of the prior and likelihood is not simple. This often happens when for example the parameter space is high dimensional which requires calculating multidimensional likelihoods. The most popular approach to approximating such integrals has been to apply numerical techniques such as importance sampling and thermodynamic integration but these confront the problem that the prior parameters are distributed in regions where the likelihood function is not highly concentrated [2,10,16]. Other recent sampling techniques for example TMCMC in [4] that are based on MCMC paradigm can be used but these tend to only sample the peaks of the posterior distribution which can under sample most of the narrow best-fit regions.

Recently Skilling in [18] proposed an algorithm, nested sampling, which is able to estimate integrals of the form shown in equation 9 efficiently. The algorithm works by transforming the multidimensional parameter space integral to a one dimensional one where classical numerical approximation techniques of estimating the area under the function can be applied. The algorithm has been successfully applied and further modified in a number of recent astronomy and cosmology papers [7, 14, 15, 20]

## 4. Sampling

4.1 Nested sampling

Nested sampling is a Monte Carlo but not a Markov Chain method of sampling. The main idea behind the method is to divide the prior parameter space into 'equal mass' units and to order these by model likelihood. The total prior mass is denoted $X$ and each unit in this prior mass is $P(\theta|H)\,d\theta = \pi(\theta)d\theta = dX$. The likelihood function is written $P(D|\theta,H) = L(\theta) = L(X)$ in this space. This formulation transforms the likelihood into a function of a one-dimension parameter. The evidence integral can now be written:

$$evidence = Z = \int L(\theta)\,\pi(\theta)\,d\theta = \int_0^1 L\,dX \tag{10}$$

The algorithm then supposes likelihoods can be evaluated at all $X_i$ so that $L_i = L(X_i)$, where $X_i$ is a sequence of values that decrease from 1 to 0 such that $0 < X_N < ..... < X_2 < X_1 < X_0 = 1$ as illustrated in the bottom image of figure 1. The top image shows a number of samples and the likelihood iso-contours in the 2-dimensional posterior probability parameter space.

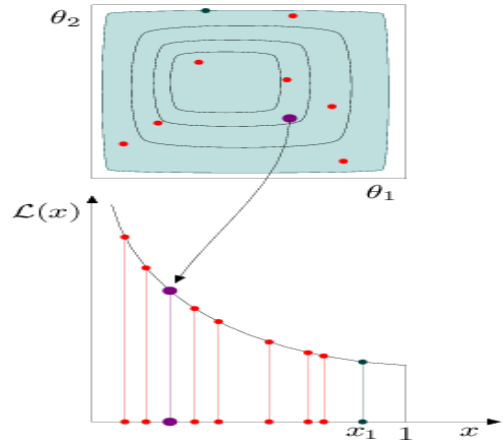

**Figure 1**: Sampling from a 2D parameter space by nested sampling

Such a one dimensional integral function, equation 10, can easily be estimated by any numerical method such as the trapezoid rule:

$$Z = \sum_{i=1}^{N} Z_i \qquad Z_i = \sum_{i=1}^{N} L_i b_i \qquad (11)$$

where $b_i = \frac{1}{2}(X_{i-1} - X_{i+1})$ and $L_i$ is the likelihood at that sample. In the context of finite element updating, the algorithm achieves the above approximation in the following manner:

1. Sample *N* updating parameters from the prior probability distribution. Evaluate their likelihoods.
2. From the *N* samples select the sample with the lowest likelihood ($L_i$).
3. Increment the evidence by $Z_i = \frac{L_i}{2}(X_{i-1} - X_{i+1})$.
4. Discard the sample with the lowest likelihood and replace it with a new point from within the remaining prior volume $[0, X_i]$. The new sample must satisfy the constraint $L_{new} > L_i$
5. Repeat steps 2-4 until some stopping criterion is reached. This could be the desired precision on the evidence or some iteration count (*Max* in our algorithm).

For further details on nested sampling see [18]. The next section presents the experiments performed on different finite element models and the evaluation of their evidences.

4.2  Model definition and comparison

In FEM a mathematical model ($H_i$ in section 3) is defined by the quantity and location of free-parameters (updating parameters). Here finite element models were designed to have a different *number* of free-parameters at different *positions* along the same structure, effectively creating different mathematical models for the same structure. In mathematics the number of free-parameters is related to the complexity of a mathematical model and to advocate Occam's razor; a model with fewer parameters is preferred. This is useful in not only determining which parameters need to be updated but also how many can be deemed sufficient to approximate the measured data well. To determine how well different models compare in fitting the data, Jeffrey's scale [10], shown in table 1, was used.

**Table 1:** Jeffrey's factor scale

| $Z_j/Z_i$ | $\log_2(Z_j/Z_i)$ | $\log_e(Z_j/Z_i)$ | $\log_{10}(Z_j/Z_i)$ | Evidence against model $H_i$ |
|---|---|---|---|---|
| 1 to 3.2 | 0 to 1.7 | 0 to 1.2 | 0 to 0.5 | Weak |
| 3.2 to 10 | 1.7 to 3.3 | 1.2 to 2.3 | 0.5 to 1 | Substantial |
| 10 to 100 | 3.3 to 6.6 | 2.3 to 4.6 | 1 to 2 | Strong |
| > 100 | > 6.6 | > 4.6 | > 2 | Decisive |
| > 1000 | > 10 | > 7 | >3 | Beyond reasonable doubt |

## 5. Experiments

A simple unsymmetrical H-beam, with six degrees of freedom, previously used in [13], is modeled. The measured natural frequencies of this structure occur at; 53.9Hz, 117.3Hz, 208.4Hz, 254Hz and 445Hz which correspond to modes 7, 8, 10, 11 and 13 respectively. To validate that model evidence calculation can reveal the most plausible finite element model(s), four *randomly* designed models of one beam structure are developed and the evidence of each is calculated. The example assumes no a prior knowledge of which updating parameters should be chosen is available. The objective is then to determine from evidence ratios, the most probable model from this random set.

The structure was modeled using the SDT® version 6.0 Matlab® toolbox. Each model used standard isotropic material properties and Euler Bernoulli beam elements to approximate the beam sections of the structure. It is assumed that the Young's modulus of some elements is not certain and was thus considered a variable to be updated in the particular updating parameters.

5.1 Unsymmetrical H beam

Figures 2 and 3 illustrate a sketch of a 600mm long unsymmetrical aluminium structure with the left vertical beam of 400mm and the right beam of 200mm. It is divided into 12 elements. Each cross-sectional area is 9.8mm by 32.2 mm (see [13] for more details on the structure and experimental set-up). The structure's elements are numbered to clarify the reading of table 2, where two sets of models are shown; model 1[A-C] and model 2 [A-C].

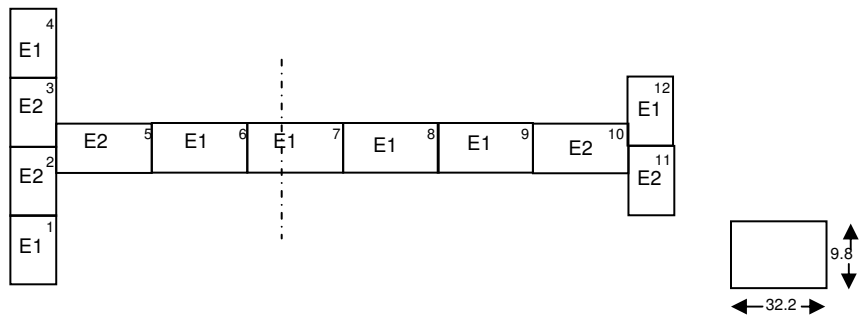

**Figure 2**: Model 1A of a 12 Element Unsymmetrical H beam.

The updating parameters were sampled from a Gaussian prior probability distribution with a mean of $7.2 \times 10^{10}$ $N.m^{-2}$ and an exponential inverse variance of 4.0e-20 for Young's modulus values between $6.8 \times 10^{10}$ $N.m^{-2}$ and $8.0 \times 10^{10}$ $N.m^{-2}$ for aluminium. The number of samples, *N*, was set to 100 (to sample the distribution well) and the sampling algorithm stopping criterion is experimentally set to a maximum of 250 iterations. Model 1A in table 2 is the modeled structure such that the elements labeled E1 in figure 2 are set to a constant value of Young's modulus for aluminium ($7.2 \times 10^{10}$ $N.m^{-2}$). The Young's modulus values for elements labeled E2 are sampled from a prior with a Gaussian probability distribution as mentioned above. Model 1B is the opposite setting where now the E2 labeled parameters are fixed and the E1 parameters are sampled from a Gaussian prior distribution. All the elements that are free or updated in each model are listed under the heading "Parameter label" in table 2. The set of updated parameters does not necessarily have to be fixed; all parameters can be varied simultaneously as shown by models 1C and 2C in table 2. Figure 3 illustrates model 2A from the second set of models.

**Table 2:** Updated parameters and model evidences

| Model | Log(Evidence) | Quantity of parameters | Parameter label |
|---|---|---|---|
| 1A | $-2.188 \pm 0.04$ | 5 | 2,3,5,10,11 |
| 1B | $-11.40 \pm 0.12$ | 7 | 1,4,6,7,8,9,12 |
| 1C | $-12.15 \pm 0.13$ | 12 | All |
| 2A | $-14.95 \pm 0.13$ | 7 | 1,5,6,7,8,9,10 |
| 2B | $-2.188 \pm 0.04$ | 5 | 2,3,4,11,12 |
| 2C | $-21.74 \pm 0.14$ | 12 | All |

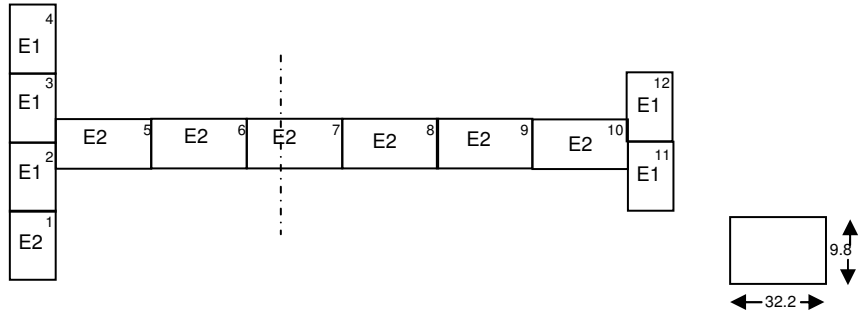

**Figure 3**: Model 2A of a 12 Element Unsymmetrical H beam.

Table 3 shows the results using Jeffrey's scale. Models 1A and 2A are very different; model 1A is a better model for this data than model 2A. It is interesting to note that the models with the least number of updating parameters produced the best evidences and this is because it is relatively simple and in accordance with Occam's razor which states that the least complex model that best describes the data is the correct one. Model 1A and 2B are shown to be relatively similar while there is strong evidence against model 2A from model 1B. So evidence calculation can provide a mechanism for eliminating poor models from the onset. It also provides a platform to determine salient parameters to consider in the updating process.

**Table 3:** Bayes factors for Unsymmetrical beam model

| $H_p / H_q$ against model $H_q$ | Bayes Factor $(Z_p / Z_q)$ | Jeffrey's Scale $\log_e(Z_j / Z_i)$ | Evidence |
|---|---|---|---|
| 1A/1B | 10122 | 9.2 | Beyond reasonable doubt |
| 1A/2A | 349760 | 12.7 | Beyond reasonable doubt |
| 1A/2B | 1 | 0 | Weak |
| 1B/ 2A | 35 | 3.5 | Strong |

## 6. Conclusion

In this paper we have introduced the model selection concept to the problem of finite element model updating. It was argued that plausible model evidences should be calculated before models can be updated. A recently proposed method of comparing mathematical models was introduced and implemented in the finite element model context. This algorithm, nested sampling, efficiently calculates the Bayesian evidence of a model and provides the posterior probability distribution of the model parameters as a by-product. A simple beam structure with a number of random mathematical model formulations, defined by the number and position of updating parameters, is used as proof-of-concept for the algorithm in this domain. Clear distinctions between the models are evident from their relative evidences.

# 7. Acknowledgements

This research was supported by the financial assistance of the National Research Foundation of South Africa. I would like to thank members of the Wits Computational Intelligence group for their helpful discussions.